\documentclass[epj,referee]{svjour}
\usepackage{graphics}
\begin{document}
\title{Entangling single photons on a beamsplitter}

\author{M. Larqu\'e, A. Beveratos \and I. Robert-Philip}                     
\offprints{isabelle.robert@lpn.cnrs.fr}          
\institute{CNRS - Laboratoire de Photonique et Nanostructures, Route de Nozay, F-91460 Marcoussis, FRANCE}
\date{Received: date / Revised version: date}
%
\abstract{
We report on a scheme for the creation of time-bin entangled states out of two subsequent single photons. Both photons arrive on the same input port of a beamsplitter and the situation in which the photons leave the beamsplitter on different output ports is post-selected. We derive a full quantum mechanical analysis of such time-bin entanglement for emitters subject to uncorrelated dephasing processes and apply this model to sequential single photons emerging from a single semiconductor quantum dot. Our results indicate that  the visibility of entanglement is degraded by decoherence effects in the quantum dot, but can be restored by use of CQED effects, namely the Purcell effect.
} 
\maketitle
\onecolumn

\section{Introduction}
\label{intro}

Since the first realization of an entangled photon source allowing for the violation of Bell inequalities \cite{Aspect1982}, entangled photons have become an important element in the arsenal of quantum optics experimental techniques and were subsequently used in many studies of fundamental aspects of quantum mechanics. More recently, it has become an enabling resource of potentially disruptive quantum information science, such as quantum communications allowing for quantum key distribution links \cite{Ekert1991} \cite{Bennett2000} \cite{Jen_Ent-Crypto_2000} or the realization of quantum relays \cite{Collins2005}. Up to now, entanglement has been realized mostly by use of two photons emitted by nonlinear optical process (such as parametric down conversion \cite{Franson1989} \cite{Kwiat_BBO_95} \cite{Tanzilli2002}). Such non linear sources are routinely used in quantum optics, particularly in experiments on quantum teleportation \cite{Pan1998} \cite{Ursin_Tele-Danube_2004} \cite{Landry_Telet_PlainPalais_07} or entanglement swapping \cite{Yang_Syncro_Indep_06} \cite{Halder2007}. However, such sources are based on parametric downconvertion, which follows a Poissonian emission statistics. On the other hand, deterministic sources of entangled photons can be obtained from cascade emission in single emitters. Polarization entangled photons have already been produced from single atoms \cite{Aspect1982} or single quantum dots \cite{Stevenson2006} \cite{Akopian2006}. A scheme producing deterministic time-bin entangled photons using the cascade emission of a single quantum dot excited from a still unknown metastable state has also been recently proposed \cite{Simon2005}. In such two photon emitters, photons emerge from a common source and are produced in a entangled state by the emission processes itself. Alternatively, deterministic polarization entangled photons can be obtained with two indistinguishable single photons emitted sequentially and linear optic components \cite{Fattal2004}. 

\begin{figure}
\begin{center}
\resizebox{0.75\columnwidth}{!}{%
  \includegraphics{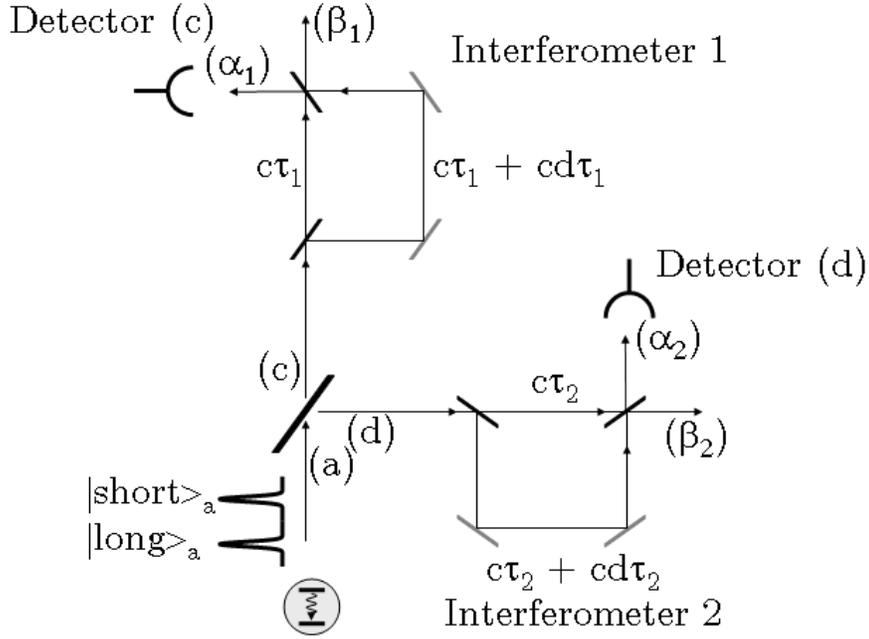}
}
\caption{Schematic of entanglement generation and analysis. Two sequential indistinguishable single photons, with a relative time delay of $T$,  are split by a beamsplitter and sent to unbalanced Mach-Zender interferometers.}
\label{fig:schematic}
\end{center}
\end{figure}

In this paper, we propose a scheme for entangling two sequential single photons on a simple beamsplitter, creating hence a deterministic time-bin entangled state \cite{Brendel1999}. Two sequential indistinguishable single photons (see Fig. \ref{fig:schematic}), separated by a time delay $T$ large compared with the single photon pulse duration, are incident on the same input port of a beamsplitter. Let us denote them $|short\rangle_a$
and $|long\rangle_a$. The two output ports of the beamsplitter are denoted $c$ and $d$. If we discard the events when both photons follow the same output port (states $|short\rangle_{c}|long\rangle_c$ and $|short\rangle_d|long\rangle_d$), the post-selected state, obtained with a probability $1/2$ at the output of the beamsplitter, reads:
\begin{equation}
|\Psi^+\rangle= \frac{1}{\sqrt{2}}(|short\rangle_c|long\rangle_d+|long\rangle_c|short\rangle_d)
\end{equation}
\noindent This state is a time-bin entangled state, that can be further analyzed in a Franson-type photon correlation set-up \cite{Franson1989} \cite{Marcikic2002}, composed of two interferometers $1$ and $2$ respectively placed along the output ports $c$ and $d$ of the beamsplitter. Postselection of state $|\Psi^+\rangle$ is achieved by recording only coincidence events between the output ports of interferometer $1$ and the output ports of interferometer $2$. In such an experiment, we perform an analysis of the entangled state created on the beamsplitter by which-path interferometry. Such a scheme relies on two crucial features, namely the efficient generation of deterministic single photon states and the indistinguishability of these states. 

Initially indistinguishable single photons were obtained from a process of degenerate parametric downconversion \cite{Hong1987} \cite{deRiedmatten2003} \cite{Kaltenbaek2006}; however, these sources based on non-linear processes are not deterministic, since the generation of photon pairs per excitation cycle is probabilistic, entailing multi-pair events that lead inevitably to a decreased degree of entanglement \cite{Scarani2005}. In recent years, the use of single photon emitters to produce quantum mechanically indistinguishable single photon sources has made possible to eliminate multiphoton occurrences. Such single emitting dipoles are for example self-assembled quantum dots \cite{Santori2002} \cite{Varoutsis2005} \cite{Laurent2005}. However, such solid-state emitters are subject to strong dephasing mechanisms, which partially degrade the degree of indistinguishability between the emitted single photons.

In this paper, we present the theory of time-bin entanglement generation from a single photon emitter by considering explicitly the effect of dephasing processes. The paper is organized as follows: In Section 2 we describe the main ingredient entering in our model,
namely, the single photon emitter subject to random phase fluctuations during emission. In Section 3, we calculate the probability of simultaneous detection between the output ports of two Mach-Zender interferometers placed on each output port of the beamsplitter and derive the visibility of the two-photon interference. These calculations serve in Section 4 as a guide for identifying experimental situations in which time-bin entanglement can be realized efficiently. Finally, in Section 5 we summarize and conclude.

\section{Single photons with phase diffusion}

We consider the situation in which single photons originate from a single-photon emitter subject to phase diffusion. The single photon emitter is modeled as a two-level system that interacts with the electromagnetic field
by absorbing or emitting photons whenever it undergoes a transition between its two states. The emission of a photon at time $t_0$ and its propagation during a time $\tau$ can be described by a creation operator $\hat{\gamma}^{(+)}_{t_0}$:

\begin{equation}
\hat{\gamma}^{(+)}_{t_0} = \int_{R} \mu(t'-t_0) \hat{a}^{(+)}(t'+\tau) dt'
\label{eq:PhotonEmission}
\end{equation} 

\noindent where $\hat{a}^{(+)}(t)$ is the photon creation operator at time $t$ and $\mu(t)$ represents the temporal shape of the single photon emitted by the two-level system. For a perfect two-level system isolated from the environment, the function $\mu(t)$ would be an exponential decay with a characteristic time constant related to the radiative lifetime $T_1=1/\Gamma'$ of the two-level system placed in its excited state. However, most single emitting dipoles are subject to sudden, brief, and random fluctuations of their energy (arising, for example, from collisions with phonons and electrostatic interactions with fluctuating charges located in the dipole vicinity \cite{Berthelot2006}). These fluctuations of the two-level system are assumed to be completely uncorrelated between them. In this context, the wavefunction $\mu(t)$ reads \cite{Bylander2003}:

\begin{equation}
\mu(t) = K e^{-i\Omega_0 t - i \phi(t) - \frac{\Gamma'}{2}t - i\Delta t} H(t)
\label{eq:wavefunction}
\end{equation}

\noindent where $K$ is a normalization constant, $H(t)$ is the Heaviside function, $\hbar \Omega_0$ is the energy of the transition energy of the two-level system isolated from its environment and $\Delta$ is the radiative frequency shift due to the interaction with the environnement. In the following, we shall denote $\Omega = \Omega_0 + \Delta$. The phase $\Phi(t)$ is the fluctuating phase of the two-level system subject to phase diffusion, which, in the interaction picture, is transferred to the emitted photon. It satisfies the following relations:

\begin{eqnarray}
\overline{ e^{i\Phi(t)}} &=& 1\\
\overline{  e^{i\Phi(t) - i\Phi(t')}} &=& e^{-\Gamma |t-t'|}
\end{eqnarray}

\noindent where the overlines denote statistical averaging. $\Gamma$ is the dephasing rate of the two-level system and can be expressed as a function of the characteristic time for pure dephasing according to $T_2^* = 1/\Gamma$. The dephasing rate $\Gamma$ and the decay rate $\Gamma'$ are related to the coherence time $T_2$ as follows:

\begin{equation}
\frac{1}{T_2} = \Gamma + \frac{\Gamma'}{2}=\frac{1}{T_2^*}+\frac{1}{2T_1}
\end{equation}

\noindent Let's suppose that after a propagation time $\tau$, the photon emitted at time $t_0$ is detected by a detector that converts the single photon received into an electric pulse that can subsequently be processed. The probability $p(t)$ of photodetection of the single photon at time $t$ can then be expressed by use of the $\hat{E}^{(-)}(t)$ photodetection annihilation operator at time $t$ as:

\begin{equation}
p(t)=\overline{\langle 0|\hat{\gamma}^{(-)}_{t_0}\hat{E}^{(+)}(t)\hat{E}^{(-)}(t)\hat{\gamma}^{(+)}_{t_0}|0\rangle}
\end{equation}

\noindent where the angular brackets denote quantum mechanical averaging with respect
to the state of the electromagnetic field and the overline the statistical averaging with respect to the fluctuations. By use of eq. (\ref{eq:PhotonEmission}) and (\ref{eq:wavefunction}), this probability reads:

\begin{eqnarray}
p(t)&=&\overline{|\mu(t-\tau-t_0)|^{2}}\\
&=&|K|^{2}e^{-\Gamma'(t-\tau-t_0)}H(t-\tau-t_0)
\end{eqnarray}

\noindent The normalization condition $\int_{\mathbf{R}}p(t)dt=1$ (no transmition losses and perfect detector) implies $|K|^2=\Gamma'$.\\

\section{Entanglement on a beamsplitter}

We suppose that the two photons, involved in the experiment described on Fig. \ref{fig:schematic}, are produced sequentially by the same two-level system, with a time delay $T>>T_1$. Both photons impinge the beamsplitter on the same input port $a$. In the following, we shall only consider the situation in which the two photons are separated by the beamsplitter along two different output ports, noted $c$ and $d$. The creation operator of the photon pair can then be expressed as:

\begin{equation}
\hat{\gamma}^{a,(+)}_{t=0}\hat{\gamma}^{a,(+)}_{t=T}{\rightarrow} \sqrt{R_{BS}T_{BS}}(\hat{\gamma}^{c,(+)}_{0}\hat{\gamma}^{d,(+)}_{T}+\hat{\gamma}^{d,(+)}_{0}\hat{\gamma}^{c,(+)}_{T})
\label{eq:gammaBS}
\end{equation}

\noindent where $\hat{\gamma}^{x,(+)}_{t_0}$ is the creation operator of a photon emitted at time $t_0$ on port $x$ of the beamsplitter. $R_{BS}$ and $T_{BS}$ are the intensity reflection and transimission coefficients of the beamsplitter. These photons propagate respectively through two unbalanced interferometers $1$ and $2$ and are further detected on two detectors placed on the outputs $\alpha_i$ ($i=1,\ 2$) of these interferometers. This setup postselects the configuration in which the photons excite different output ports of the beamsplitter. In addition, joint photodetections are registered only for delays between detection events lower than the delay $T$ between photon emission occurences. The results of such an experiment give the probability $p_{12}$ of joint photodetection of one photon on output port $\alpha_1$ and of one photon on output port $\alpha_2$, provided the first (resp. second) photon passed through the long (resp. short) arm of the interferometer. This probability is related to the interference between two possible paths leading to photons on outputs $\alpha_1$ and $\alpha_2$, with no information (in ideal case) on which photon was the first or the second emitted by the source. The other joint photodetection events with delay greater than $T$ are not resulting from this interference but can be used experimentally to normalize the probability $p_{12}$.\\

\subsection{One photon travelling through one interferometer}

\begin{figure}
\begin{center}
\resizebox{0.75\columnwidth}{!}{%
  \includegraphics{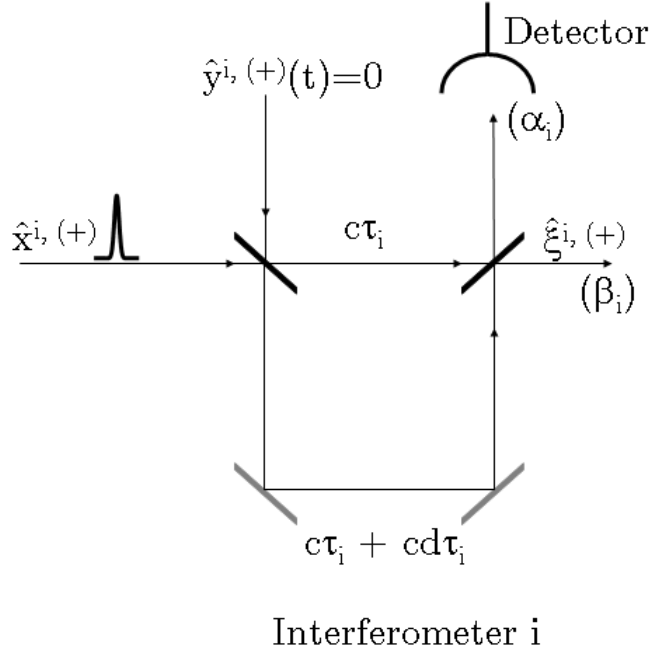}
}
\caption{Schematic of an unbalanced interferometer.}
\label{fig:interferometer}\end{center}
\end{figure}

In order to express $p_{12}$, let us first consider the situation in which one single photon emitted at time $t_0$ travels through an unbalanced interferometer with arm lengths equal to $c\tau_i$ and $c(\tau_i+d\tau_i)$ as described on Fig. \ref{fig:interferometer}. We suppose that one input state of the interferometer is vacuum ($\hat{y}^{i, (+)}(t)=0$) and that the other input state $\hat{x}^{i, (+)}(t)$ is the single photon state arriving at time $t$ on the input port. The creation operator $\hat{x}^{i,(+)}(t)$ and the operator $\hat{\xi}^{i,(+)}(t')$ of creation at time $t'$ of one photon along the output $\alpha_i$ of the interferometer are related by:
\begin{equation}
\hat{x}^{i, (+)}(t) = \sqrt{R_iT_i} \hat{\xi}^{i, (+)}(t+\tau_i+d\tau_i) -\sqrt{R_iT_i} \hat{\xi}^{i, (+)}(t+\tau_i)
\label{eq:TransmissionInterfero}
\end{equation}

\noindent The first term in the previous equation corresponds to the situation in which the photon has followed the long arm of the interferometer, while the second term corresponds to a propagation along the short arm. $R_{i}$ and $T_{i}$ are the intensity reflection and transimission coefficients of the interferometer. In this context, the detection probability $p_i$ of a single photon along the $\alpha_i$ output of the interferometer at time $t$ reads:

\begin{equation}
p_i(t)=\overline{\langle 0|(\hat{\alpha}^{li, (-)}_{t_0}+\hat{\alpha}^{si,(-)}_{t_0})\hat{E}_i^{(+)}(t)\hat{E}_i^{(-)}(t)(\hat{\alpha}^{li, (+)}_{t_0}+\hat{\alpha}^{si, (+)}_{t_0})|0\rangle}= \overline{ \left\| \hat{E}_i^{(-)}(t)(\hat{\alpha}^{li, (+)}_{t_0}+\hat{\alpha}^{si, (+)}_{t_0})|0\rangle \right\|^2}
\end{equation}

\noindent where $\hat{E}_i^{(-)}(t)$ is the photodetection annihilation operator at time $t$ on the detector placed on the $\alpha_i$ output. The creation operators $\hat{\alpha}^{li, (+)}_{t_0}$ and $\hat{\alpha}^{si, (+)}_{t_0}$ can be expressed as follows by use of equations (\ref{eq:PhotonEmission}) and (\ref{eq:TransmissionInterfero}):

\begin{equation}
\hat{\alpha}^{li, (+)}_{t_0} = \sqrt{R_iT_i}\int_{R} \mu(t'-t_0)\hat{\xi}^{i, (+)}(t+\tau_i+d\tau_i)dt'
\label{eq:alphali}
\end{equation}

\begin{equation}
\hat{\alpha}^{si, (+)}_{t_0}  = -\sqrt{R_iT_i}\int_{R} \mu(t'-t_0) \hat{\xi}^{i, (+)}(t+\tau_i) dt'
\label{eq:alphasi}
\end{equation}

\noindent In these expressions, only propagation times through the interferometers are taken into account. 

\subsection{Two photons travelling through two different interferometers}

In the experiment involving two photons separated by a time delay $T$, joint photodetection can only occur if the single photon state created at time $t=0$ follows the long arm of one interferometer, and the single photon state created at time $t=T$ follows the short arm of the other interferometer. In this context, the probability $p_{12}(t_1, t_2)$ of photodetection of one single photon on $\alpha_1$ port at time $t_1$ and of one single photon on $\alpha_2$ port at time $t_2$ reduces, by use of (\ref{eq:gammaBS}), to:

\begin{equation}
p_{12}(t_1, t_2)=R_{BS}T_{BS} \overline{ \left\| \hat{E}_2^{(-)}(t_2)\hat{E}_1^{(-)}(t_1)(\hat{\alpha}^{l1, (+)}_{0}\hat{\alpha}^{s2, (+)}_{T}+\hat{\alpha}^{l2, (+)}_{0}\hat{\alpha}^{s1, (+)}_{T})|0\rangle \right\|^2 }
\label{eq:p12t1t2}
\end{equation}

\noindent Using relations (\ref{eq:alphali}) and (\ref{eq:alphasi}), the joint photodetection probability $p_{12}$ can be evaluated as:

\begin{eqnarray}
p_{12}&=& \int_{\mathbf{R}^2}dt_1 dt_2 p_{12}(t_1, t_2)\\
&=& R_{BS}T_{BS}R_1T_1R_2T_2  \int_{\mathbf{R}^6} \prod_{i=1}^6 dt_i \ \overline{\mu(t_3)^* \mu(t_4-T)^* \mu(t_5-T) \mu(t_6)} \nonumber\\ 
&& 
\langle 0| [\hat{\xi}^{1, (-)}(t_3+\tau_1+d\tau_1) \hat{\xi}^{2, (-)}(t_4+\tau_2) + \hat{\xi}^{2, (-)}(t_3+\tau_2+d\tau_2) \hat{\xi}^{1, (-)}(t_4+\tau_1)]\hat{E}_1^{(+)}(t_1)\hat{E}_2^{(+)}(t_2) \nonumber\\ 
&& 
\hat{E}_2^{(-)}(t_2)\hat{E}_1^{(-)}(t_1) [\hat{\xi}^{1, (+)}(t_6+\tau_1+d\tau_1) \hat{\xi}^{2, (+)}(t_5+\tau_2) + \hat{\xi}^{2, (+)}(t_6+\tau_2+d\tau_2) \hat{\xi}^{1, (+)}(t_5+\tau_1)]|0\rangle\nonumber\\
\end{eqnarray}

\noindent In this expression, integration over all detection times $t_1$ and $t_2$ and not only for $|t_2-t_1|<T$ is allowed, since only the paths leading to joint photodetection are selected in equation (\ref{eq:p12t1t2}). This integral can be readily evaluated by commuting the two annihilation (resp. creation) electric field operators through the photon creation (resp. annihilation) operators. Operating these commutations, the joint photodetection probability can be written as:

\begin{eqnarray}
p_{12}&=& R_{BS}T_{BS}R_1T_1R_2T_2  \int_{\mathbf{R}^6} \prod_{i=1}^6 dt_i \ \overline{\mu(t_3)^* \mu(t_4-T)^* \mu(t_5-T) \mu(t_6)} \nonumber\\ 
&& 
[\delta(t_3+\tau_1+d\tau_1-t_1) \delta(t_4+\tau_2-t_2) + \delta(t_3+\tau_2+d\tau_2-t_2) \delta(t_4+\tau_1-t_1)] \nonumber\\ 
&& 
[\delta(t_6+\tau_1+d\tau_1-t_1) \delta(t_5+\tau_2-t_2) + \delta(t_6+\tau_2+d\tau_2-t_2) \delta(t_5+\tau_1-t_1)]\\
&=& R_{BS}T_{BS}R_1T_1R_2T_2  \int_{\mathbf{R}^2} dt_1 dt_2 \nonumber\\
&& \overline{\mu(t_1-\tau_1-d\tau_1)^2}\ \overline{\mu(t_2-\tau_2-T)^2} + \overline{\mu(t_2-\tau_2-d\tau_2)^2}\ \overline{\mu(t_1-\tau_1-T)^2}\nonumber\\
&& + (\overline{\mu(t_1-\tau_1-d\tau_1)^*\mu(t_2-\tau_2-d\tau_2)}\ \overline{\mu(t_2-\tau_2-T)^*\mu(t_1-\tau_1-T)} + c.c.)
\end{eqnarray}

\noindent where ``c.c.'' denotes the complex conjugate. The emission process by the source is supposed to be much faster than the delay $T$ between the two emission events, so that the emission of the first photon does not affect the emission of the second one, and statistical averaging can be made independently on each photon. Thus, using the expression of $\mu(t)$ given by equation (\ref{eq:wavefunction}) and upon integration, we obtain the probability of joint photodetection of single photons on both outputs $\alpha_1$ and $\alpha_2$ as:

\begin{equation}
p_{12}= 2R_{BS}T_{BS}R_1T_1R_2T_2 [1+ V\times cos(\Omega(d\tau_1-d\tau_2))]
\label{eq:ProbabiliteDeCoincidenceFinale}
\end{equation}

\noindent where:
\begin{equation}
V = e^{-\frac{\Gamma'}{2}(|T-d\tau_1|+|T-d\tau_2|)-\Gamma |d\tau_1-d\tau_2|} [1+e^{-\frac{\Gamma'}{2}|d\tau_1-d\tau_2|}cosh(\Gamma'\frac{|T-d\tau_1| - |T-d\tau_2|}{2})(\frac{T_2}{2T_1}-1)]
\label{eq:VisibilityFinale}
\end{equation}

The coincidence rate between the two detectors displays an oscillatory behavior as a function of the phase difference $\Omega(d\tau_1-d\tau_2)$ between the two interferometers. $V$ is the visibility of the interference fringes. This interference is a two-photon interference and results from equal probability amplitudes of two indistinguishable paths: (1) first and second photons respectively in the long arm of interferometer $1$ and short arm of interferometer $2$ and (2) first and second photons respectively in the long arm of interferometer $2$ and short arm of interferometer $1$. As expected, the interference pattern only depends on the phase difference of the two analyzing interferometers. Contrary to experiments with time-bin entangled parametric down converted photons \cite{Marcikic2002}, it does not depend on the phase of the pump interferometer, since the photon emission is an incoherent effect.

\section{Restoration of Entanglement through the Purcell effect}

The visibility $V$ is experimentally measured as a function of the optical path difference $c(d\tau_1-d\tau_2)$ between the two interferometers. It falls with increasing time delay $(d\tau_1-d\tau_2)$: this decrease is related to the timing information one could extract from the detection events and give rise to which-path information. However, $V$ is a slow-varying function of $(d\tau_1-d\tau_2)$ compared with the cosine term. Therefore, its maximum amplitude can be inferred from experiments in which $d\tau_1 \approx d\tau_2 \approx T$ within small variations of $(d\tau_1-d\tau_2)$ compared with $1/\Gamma$ and $1/\Gamma'$.

Perfect visibility $(V=1)$ is obtained for perfectly indistinguishable single photons ($T_2=2T_1$). In the absence of pure dephasing, the produced entangled state is maximally entangled, hence giving rise to perfect correlations. On the other hand, the maximal visibility drops to $T_2/(2T_1)$ for single photons which are not Fourier-transform limited ($T_2<2T_1$). This value of $T_2/(2T_1)$ is the ``degree of indistinguishability'' of the single photons, i.e. the amplitude of the coincidence dip observed in a Hong-Ou-Mandel two photon interferometer, which implements the coalescence of two single photons impinging on different input ports of a beamsplitter \cite{Hong1987}. Yet, a visibility lower than unity does not imply that the state is not entangled. Among the different criteria demonstrating entanglement, one of them is the violation of Clauser-Horne-Shimony-Holt (CHSH) inequality. The violation of such inequality involves a visibility $V$ higher than $1/\sqrt{2}$ \cite{Tittel1999}.

In order to illustrate our results with a concrete example, we shall consider the case of a semiconductor quantum dot as the single emitting dipole. Their lifetime is generally equal to $T_1=1$ ns, and the pure dephasing time varies from $T_2^*=30$ ps to $T_2^*=300$ ps and even values higher than 1 ns for quantum dots under non-resonant pumping in the wetting layer, quasi-resonant pumping on a confined excited state \cite{Kammerer2002} and under resonant pumping \cite{Borri2001} \cite{Langbein2004} respectively. These variations of the pure dephasing time on the pumping conditions can be explained by dephasing electrostatic interactions of the trapped exciton with fluctuating charges located in the dipole vicinity \cite{Berthelot2006} and values higher than 1 ns can only be reached under resonant pumping \cite{Borri2001} \cite{Langbein2004}, which is precluded for single photon generation. For pure dephasing times respectively equal to 30 ps, 300 ps and 1 ns, the visibility given from equation (\ref{eq:VisibilityFinale}) is equal to $V_{30ps}=0.014$, $V_{300ps}=0.130$ and $V_{1ns}=0.33$ respectively, but not high enough to violate Bell inequalities. Conversely, if $T_2^*$ reaches values much higher than 1 ns, it is possible to obtain a visibility higher than $\sqrt{2}$ and our model correctly predicts the recently and independently obtained experimental values \cite{Bennett2007} after correction of the multiphoton events \cite{Santori2002}.

\begin{figure}
\begin{center}
\resizebox{0.75\columnwidth}{!}{%
  \includegraphics{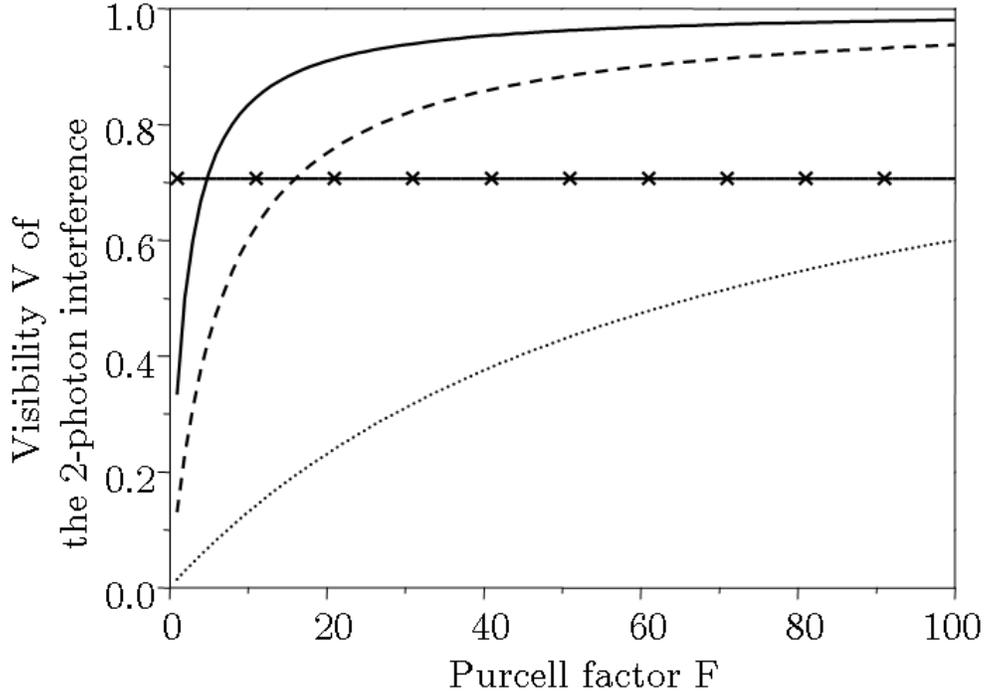}
}
\caption{Maximum visibility of the two-photon interference as a function of the Purcell factor for a two-level system displaying a spontaneous emission lifetime $T_1=1$ ns. (a) dotted line: $T_2^*=30$ ps, of the order of magnitude of the pure dephasing time of quantum dots under non-resonant pumping in the wetting layer. (b) dash line: $T_2^*=300$ ps, of the order of magnitude of the pure dephasing time of quantum dots under quasi-resonant pumping on a confined excited state. (c) solid line: $T_2^* = 1$ ns, of the order of magnitude of the pure dephasing time of quantum dots under resonant pumping. (d) Solid line with crosses:  value of $1/\sqrt{2}$}
\label{fig:visibility}\end{center}
\end{figure}

Since pure dephasing time can not be further increased, one has to decrease the dipole's lifetime $T_1$ such as the photon is emitted faster than any dephasing mechanism. This is obtained by cavity quantum electrodynamics: the density of optical states is increased by use of an appropriate microcavity, leading to a decrease in the dipole's lifetime as predicted by Purcell. First observation of cavity quantum electrodynamics effects in semiconductor dots was obtained by Gerard \textit{et. al.} \cite{Gerard1998} and Purcell factors (i.e. spontaneous emission enhancement) as high as 10 have been observed by several groups. Such values of the Purcell effect leads to a ratio of $T_2/(2T_1)$ as high as 0.8 \cite{Santori2002} \cite{Varoutsis2005} \cite{Laurent2005}. Fig. \ref{fig:visibility} depicts the maximum visibility $V$ of the interference pattern for such emitters, as a function of the spontaneous emission enhancement $F$ for different pure dephasing time $T_2^*$ and $d\tau_1 \approx d\tau_2 \approx T$ . It confirms that improvement of the visibility is achieved with the exploitation of the Purcell effect. For quantum dots with an excitation above the wetting layer band edge however, a Purcell factor higher than 160 would be necessary (Fig \ref{fig:visibility} - dotted line). Conversely, for dots with a quasi-resonant excitation, a Purcell factor of the order of 16 would be enough to violate Bell's inequality (Fig \ref{fig:visibility} - dashed line). Such Purcell factors have already been achieved, indicating that the possibility of realizing time-bin entangled photons with semiconductor quantum dots embedded in microcavities is totally accessible with available technology.

The main advantage of this scheme compared with parametric downconversion time-bin entangled schemes resides in the fact that it is not necessary to compensate for any phase drifts of the time delay $T$ between two emission events. As long as the phase drifts $\pm dT$ are small compared with the photon pulse duration $T_1$, the interference visibility remains almost constant. Indeed, $T$ may fluctuate from one experimental realization to another, for instance if the duration of the excitation pulse is not negligible compared to the duration $T_1$ of the emitted photon \cite{deRiedmatten2003}. For $d\tau_1=d\tau_2$, derivation of equation (\ref{eq:VisibilityFinale}) indicates that:
\begin{equation}
\mbox{for $d\tau_1 = d\tau_2$: } \frac{dV}{V} = -\Gamma'|dT| = - \frac{|dT| F}{T_1^{vac}}
\end{equation}

\noindent $T_1^{vac}$ corresponds to the quantum dot radiative lifetime with no spontaneous emission enhancement. This equation indicates that any variations of $T$ within the pulses duration $T_1$ will reduce the mean overlap between the wave packets of the two photons and thus reduce the visibility of the two-photon interference. As an example, variations of an amplitude of $2dT$ of the order of 1 ps corresponds to a maximum visibility of 79.8$\%$ instead of 81$\%$ for dots with $T_1=1$ ns, $T_2^*=300$ ps and subjected to a Purcell effect with $F=30$. 

\begin{figure}
\begin{center}
\resizebox{0.75\columnwidth}{!}{%
  \includegraphics{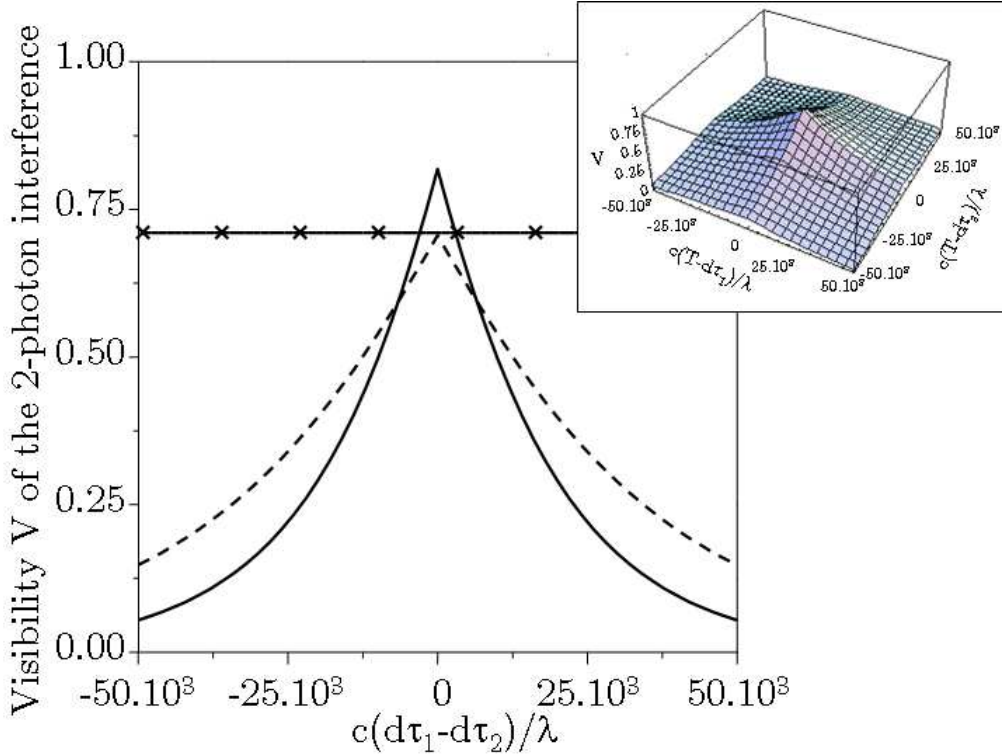}
}
\caption{Visibility $V$ of the two-photon interference as a function of the delay $d\tau_1-d\tau_2$ for $T=d\tau_1$ and for photons emerging from a single emitter with $T_1=1$ ns, $T_2^*=300$ ps and subject to the Purcell effect with: (a) dashed line: $F=16$ and (b) solid line: $F=30$. Solid line with crosses corresponds to a value of the visibility of $1/\sqrt{2}$. Insert: visibility $V$ of the two-photon interference as a function of the delay $T-d\tau_1$ and $T-d\tau_2$ for photons emerging from a single emitter with $T_1=1$ ns, $T_2^*=300$ ps and subject to the Purcell effect with $F=16$.}
\label{fig:visibility2}\end{center}
\end{figure}

In previous paragraphs, we have assumed that $d\tau_1 \approx d\tau_2 \approx T$. This can be achieved only with a certain precision. Figure \ref{fig:visibility2} shows the visibility $V$ of the two-photon interference as a function of $d\tau_1-d\tau_2$ in two situations. For two-photon interferences whose maximum visibility is equal to $1/\sqrt{2}$ (dashed line), observation of entanglement requires that the arm length differences $cd\tau_i$ of both interferometers be equal to $cT$ with wavelength-scale precision. This situation occurs for quantum dots with $T_1=1$ ns, $T_2^*=300$ ps and subject to a Purcell effect with $F=16$. Such balancing of the interferometers is experimentally not easily obtained, and interference fringes will rapidly lose visibility when scanning the interferometers. Consequently, violation of Bell inequalities requires the use of dots with a higher Purcell factor $F$. For Purcell factors as high as 30, violation of Bell inequalities is achieved if the arms length difference $c(d\tau_1-d\tau_2)$ of the two interferometers is positioned within $\pm$ 3144 $\lambda$ where $\lambda$ is the wavelength of the photon involved in the interference (see solid line on Fig. \ref{fig:visibility2}). For quantum dots emitting around 900 nm, this corresponds to a precision of $c(d\tau_1-d\tau_2)$ of the order of $\pm$ 2.8 mm achievable with usual Mach-Zender interferometer balancing techniques.

\section{Summary and conclusion}

We have studied the generation of time-bin entanglement by use of two single photons incident on a beamsplitter. These photons are be emitted sequentially with a time delay $T$ by a two-level system that undergoes dephasing in the course of the emission process. This emission process generates a state $|short\rangle + |long\rangle$ sent on a fifty-fifty beamsplitter. If we post-select the situation in which the two photons leave on two different output ports $c$ and $d$ of the beamsplitter, the obtained state after the beamsplitter is the time-bin entangled state $|\Psi^+\rangle= (|short\rangle_c|long\rangle_d+|long\rangle_c|short\rangle_d)/\sqrt{2}$.

The fidelity of entanglement is related to the visibility of the two-photon interference obtained in a Franson-type set-up composed of two interferometers. This visibility $V$ can be inferred from joint photodetection measurements between the respective outputs of the two interferometers. We have calculated the visibility $V$ for two photons originating in a single emitter characterized by an excited state lifetime $T_1 $ and a coherence time $T_2$, by modeling the dephasing process as resulting from stochastic fluctuations of the excited state energy of the emitter corresponding, for example, to fluctuations due to collisions. Our results indicate that the maximum visibility of the interference is equal to $T_2/(2T_1)$. 

Calculations using the physical parameters of semiconductor quantum dots highlight that, because of the relatively rapid dephasing, the visibility of the two-photon interference produced by bare quantum dots is very low (of the order of 10-30\%) and does not allow the violation of Bell's inequality. However, by enhancing the spontaneous emission lifetime of the quantum dots by a factor of 30 by use of Cavity Quantum Electrodynamics effects, it is possible to raise this visibility to levels that could become interesting for quantum information processing schemes. Experiments are in progress in our laboratory to implement these ideas.\\

\noindent
{\bf Acknowledgements:}
Numerous helpful discussions with I. Abram, S. Varoutsis, P. Kramper and S. Laurent
are gratefully acknowledged. This work was partly supported by the NanoSci-ERA
European Consortium under project ``NanoEPR''. We also acknowledge
support of the ``SANDiE'' Network of Excellence of the European
Commission.


\begin{thebibliography}{}


\bibitem{Aspect1982}
A.\ Aspect, P.\ Grangier and G. Roger, Phys. Rev. Lett. \textbf{49}, (1982) 91-94

\bibitem{Ekert1991}
A. Ekert, Phys. Rev. Lett. \textbf{67}, (1991) 661  

\bibitem{Bennett2000}
C. H. Bennett and D. P. DiVincenzo, Nature \textbf{404}, (2000) 247  

\bibitem{Jen_Ent-Crypto_2000}
T Jennewein, C Simon, G Weihs, H Weinfurter, A Zeilinger Phys. Rev. Lett. \textbf{84}, (2000) 4729 

\bibitem{Collins2005} D. Collins, N. Gisin, H. de Riedmatten J. Mod. Opt., \textbf{52}, (2005) 735 

\bibitem{Franson1989}
J. D. Franson, Phys. Rev. Lett. \textbf{62}, (1989) 2205 

\bibitem {Kwiat_BBO_95}
P. G. Kwiat, K. Mattle, H. Weinfurter, A. Zeilinger, A. V. Sergienko and Y. Shih Phys. Rev. Lett. \textbf{75}, (1995) 4337 

\bibitem{Tanzilli2002} S. Tanzilli, W. Tittel, H. de Riedmatten, H. Zbinden, P. Baldi, M.P. De Micheli, D.B. Ostrowsky and N. Gisin, Eur. Phys. J. D \textbf{18}, (2002) 155 

\bibitem{Stevenson2006}
R.M. Stevenson, R. J. Young, P. Atkinson, K. Cooper, D. A. Ritchie and A. J. Shields, Nature \textbf{439}, (2006) 179 
  
\bibitem{Akopian2006}
N. Akopian, N. H. Lindner, E. Poem, Y. Berlatzky, J. Avron, D. Gershoni, B. D. Gerardot and P. M. Petroff, Phys. Rev. Lett. \textbf{96}, (2006) 130501 

\bibitem{Pan1998}
J.-W. Pan, D. Bouwmeester, H. Weinfurter, and A. Zeilinger, Phys. Rev. Lett. \textbf{80}, (1998) 3891 

\bibitem{Ursin_Tele-Danube_2004}
R. Ursin, T. Jennewein, M. Aspelmeyer, R. Kaltenbaek, M. Lindenthal, A. Zeilinger Nature \textbf{430}, (2004) 849 

\bibitem{Landry_Telet_PlainPalais_07}
O. Landry, J. A. W. van Houwelingen, A. Beveratos, H. Zbinden, N. Gisin JOSA B, \textbf{24}, (2007) 398 

\bibitem{Yang_Syncro_Indep_06}
T. Yang, Q. Zhang, T.-Y. Chen, S. Lu, J. Yin, J.-W. Pan, Z.-Y. Wei, J.-R. Tian and J. Zhang Phys. Rev. Lett. \textbf{96}, (2006) 110501 

\bibitem{Halder2007} M. Halder, A. Beveratos, N. Gisin, V. Scarani, C. Simon and H. Zbinden, Nature Phys. \textbf{3}, (2007) 692 

\bibitem{Fattal2004}
D. Fattal, K. Inoue, J. Vuckovic, C. Santori, G. S. Solomon and Y. Yamamoto, Phys. Rev. Lett. \textbf{92}, (2004) 037903 

\bibitem{Simon2005}
C. Simon and J.P. Poizat, Phys. Rev. Lett. \textbf{94}, (2005) 030502 

\bibitem{Brendel1999}
J. Brendel, N. Gisin, W. Tittel and H. Zbinden, Phys. Rev. Lett. \textbf{82}, (1999) 2594 

\bibitem{Marcikic2002} I. Marcikic, H. de Riedmatten, W. Tittel, V. Scarani, H. Zbinden, N. Gisin Phys. Rev. A, \textbf{66}, (2002) 062308 

\bibitem{Hong1987}
C.K.\ Hong, Z.Y.\ Ou, and L.\ Mandel, Phys.\ Rev.\ Lett.\ \textbf{59}, (1987) 2044 

\bibitem{deRiedmatten2003} H. de Riedmatten, I. Marcikic, W. Tittel, H. Zbinden and N. Gisin
Phys. Rev. A, \textbf{67}, (2003) 022301 

\bibitem{Kaltenbaek2006} R. Kaltenbaek, B. Blauensteiner, M. Zukowski, M. Aspelmeyer and A Zeilinger Phys. Rev. Lett. \textbf{96}, (2006) 240502 

\bibitem{Scarani2005}
V. Scarani, H. de Riedmatten, I. Marcikic, H. Zbinden and N. Gisin, Eur. Phys. J. D \textbf{32}, (2005) 129 

\bibitem{Santori2002}
C.\ Santori, D.\ Fattal, J.\ Vuckovic, G.S.\ Solomon, and Y.\
Yamamoto, Nature {\bf 419}, (2002) 594 

\bibitem{Varoutsis2005}
S.\ Varoutsis, S.\ Laurent, P.\ Kramper, A.\ Lema\^itre, I.\
Sagnes, I.\ Robert-Philip, and I.\ Abram, Phys.\ Rev.\ B {\bf 72}, (2005) 041303(R) 

\bibitem{Laurent2005}
S.\ Laurent, S.\ Varoutsis, L.\ Le Gratiet,  A.\ Lema\^itre, I.\ Sagnes, F. Raineri, A. Levenson, I.\ Robert-Philip, and I.\ Abram., Appl. Phys. Lett. \textbf{87}, (2005) 163107  

\bibitem{Berthelot2006}
A. Berthelot, I. Favero, G. Cassabois, C. Voisin, C. Delalande, Ph. Roussignol, R. Ferreira and J. M. G\'erard,  Nature Physics  \textbf{2}, (2006) 759 

\bibitem{Bylander2003}
J. Bylander, I. Robert-Philip and I. Abram,  	Eur. Phys. J. D \textbf{22}, (2003) 295 

\bibitem{Tittel1999}
W. Tittel, J. Brendel, N. Gisin and H. Zbinden Phys. Rev. A, \textbf{59}, (1999) 4150 

\bibitem{Gerard1998}
J.M.\ G\'erard, B.\ Sermage, B.\ Gayral, B.\ Legrand, E.\ Costard, and V.\ Thierry-Mieg, Phys.\ Rev.\ Lett.\ \textbf{81}, (1998) 1110 

\bibitem{Kammerer2002}
C. Kammerer, C. Voisin, G. Cassabois, C. Delalande, Ph. Roussignol, F. Klopf, J. P. Reithmaier, A. Forchel and J. M. G\'erard, Phys. Rev. B \textbf{66}, (2002) 041306(R) 

\bibitem{Borri2001}
P. Borri, W. Langbein, S. Schneider, U. Woggon, R. L. Sellin, D. Ouyang and D. Bimberg, Phys. Rev. Lett. \textbf{87}, (2001) 157401 

\bibitem{Langbein2004}
W. Langbein, P. Borri, U. Woggon, V. Stavarache, D. Reuter, and A. D. Wieck Phys. Rev. B \textbf{70}, (2004) 033301 

\bibitem{Bennett2007}
A. J. Bennett, D. G. Gevaux, Z. L. Yuan, A. J. Shields, P. Atkinson and D. A. Ritchie,(2007) arxiv: 0709.0847 

\end{thebibliography}
\end{document}